\documentclass[useAMS,usenatbib]{mn2e}

\usepackage{epsfig}
\usepackage{longtable}
\usepackage{times} 
\def \apj {ApJ}

\def \apjl {ApJL}
\def \aap {A\&A}
\def \mnras {MNRAS}

\title[  ]
{An unexpected drop in the magnetic field of the X-ray pulsar
V0332+53 after the bright outburst occurred in 2015}

\author[G.\ Cusumano et al.]{G.\ Cusumano$^{1}$, V.\ La Parola$^{1}$,
 A.\ D'A\`i$^{1}$, A.\ Segreto $^{1}$, G.\ Tagliaferri$^{2}$, 
 S.D.\ Barthelmy$^{3}$, 
\newauthor N.\ Gehrels$^{3}$
\smallskip \\
$^{1}$ INAF - Istituto di Astrofisica Spaziale e Fisica Cosmica, 
    Via U.\ La Malfa 153, I-90146 Palermo, Italy\\
$^{2}$ INAF - Brera Astronomical Observatory, via Bianchi 46, 23807, Merate
    (LC), Italy\\
$^{3}$ NASA/Goddard Space Flight Center, Greenbelt, Maryland 20771, USA
\\
}

\begin{document}

\date{}

\pagerange{\pageref{firstpage}--\pageref{lastpage}} \pubyear{}

\maketitle

\label{firstpage}

\begin{abstract}
How the accreted mass settling on the surface of a
neutron star affects the
topology of the magnetic field and how  the secular evolution of the binary
system depends on the magnetic field change is still an open issue.
We report evidence for a clear drop in the observed magnetic field in 
the accreting pulsar V0332+53 after undergoing a bright 3-month long X-ray 
outburst.
We determine the field from the position of the fundamental cyclotron line in
its X-ray spectrum and relate it
to the luminosity. For equal levels of luminosity, in the declining phase 
we measure a systematically lower value of the cyclotron line energy with
respect to the rising phase. This results in a  
drop of $\sim1.7\times10^{11}$ G 
of the observed field between the onset and the end of the
outburst. The settling of the accreted plasma onto the
polar cap seems to induce a distortion of the magnetic field lines 
weakening their intensity along the accretion columns.
Therefore the dissipation rate of the magnetic field could be much faster than 
previously estimated, unless the field is able to restore its 
original configuration on a time-scale comparable with the outbursts 
recurrence time.

\end{abstract}

\begin{keywords}
X-rays: binaries -- pulsars: individual: V0332+53 -- magnetic fields -- 

\noindent
Facility: {\it Swift}

\end{keywords}


	\section{Introduction\label{intro}}
Neutron stars in high mass X-ray binaries are powered by the release of
gravitational energy from the matter accreting from their companion star. 
The magnetic field of the NS ($\sim 10^{12}$ G) drives the accreting 
matter along its field lines towards the magnetic polar caps. As a result 
an accretion column forms, where matter is slowed up by radiative processes that
produce X-rays. 
\begin{figure}
\begin{center}
\includegraphics[width=9cm]{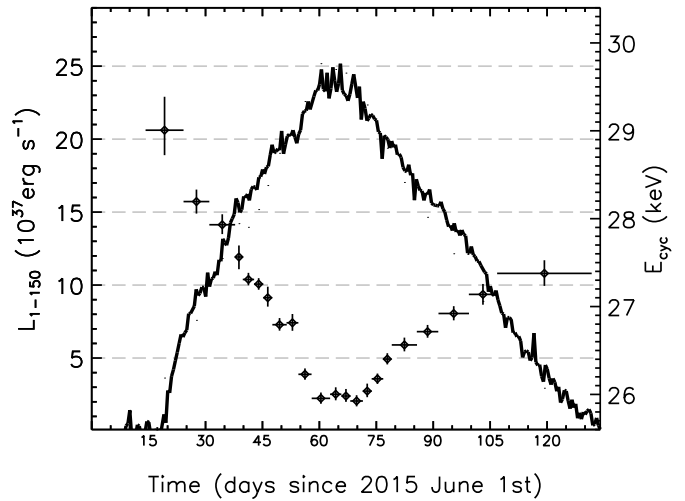}

        \end{center}
\caption{Temporal evolution of the X-ray luminosity (left axis, solid line) 
and of the energy of the CRSF fundamental line (right axis, diamond points) 
The horizontal error bars represent the time interval corresponding to 
each spectrum.
}
\label{lc}
        \end{figure}

The same magnetic field modifies the radiation emitted in the regions above the
polar caps: in the presence of such a strong magnetic field, the kinetic energy 
of the electrons in the accreting plasma is quantized in 
discrete Landau levels and photons with energies corresponding to these levels 
undergo resonant scattering, imprinting on the X-ray spectrum 
cyclotron resonant scattering features (CRSF). 
The energy of the fundamental line provides a  direct measure of 
the magnetic field in the region where the line is produced, according to the 
relation  $E_{\rm cyc}= 11.68\times B_{12} / (1+z)$ keV, where $z$ is the 
gravitational redshift in the line forming region and $B_{12}$ is the 
magnetic field in units of $10^{12}$ G.
CRSFs have been observed in $\sim20$ pulsars \citep{revnivtsev15}, probing magnetic 
fields in the range $(0.1-5)\times10^{12}$ G.
In some sources it has been observed that the position of the line changes as a 
function of the luminosity implying that the optically thick region 
where the line is imprinted moves along the accretion columns for as much as 
some hundred meters \citep{becker12}. 
 \begin{figure*}
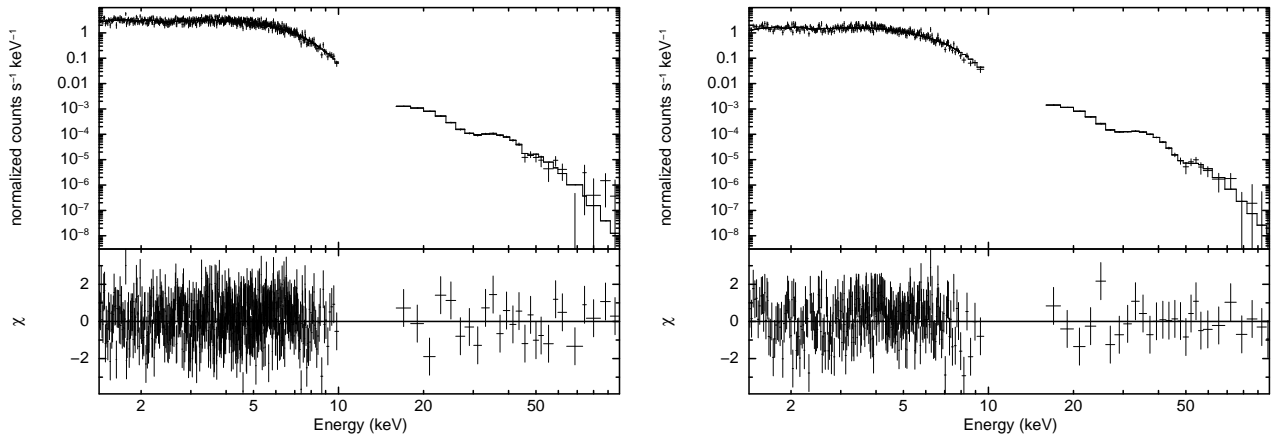

\begin{center}
\includegraphics[width=5.7cm,angle=270]{sp01_2a_fit.ps}
\includegraphics[width=5.7cm,angle=270]{sp18_fit.ps}

        \end{center}
\caption{Data, best fit model and residuals for spectra 2 (left panel) and 21 
 (right panel)}               
		\label{spec2_21} 
        \end{figure*}

\begin{figure*}
\begin{center}
\includegraphics[width=17cm]{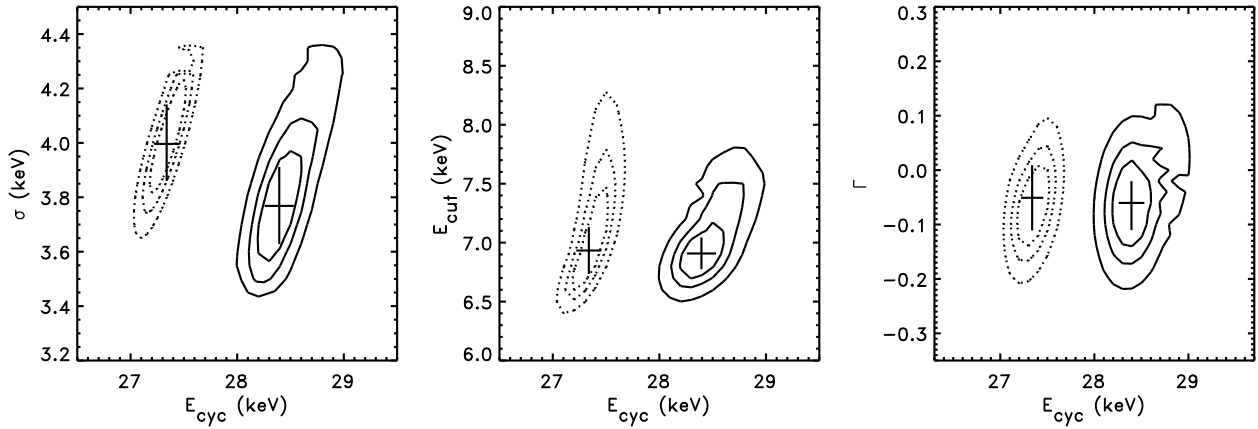}

        \end{center}
\caption{Confidence contour plots for three pairs of fit parameters, evaluated
for spectra 2 (solid line) and 21 (dotted line).  Contours are at 68\%, 90\% and
99\%. The bars indicate the 68\% confidence error ranges reported in
Table~\ref{fit}}               
		\label{contour} 
        \end{figure*}

V0332+53 is an accreting X-ray pulsar with a spin period of $\sim 4.4$ s, 
orbiting around an early type companion star \citep{negueruela99} in an 
eccentric orbit of $\sim 34$ d \citep{stella85}. The source shows
sporadic giant X-ray outbursts lasting several weeks, followed by years-long 
intervals of quiescence.  In 1989 an 
outburst recorded by the Ginga observatory revealed the presence of a cyclotron 
absorption feature \citep{makishima90} at $\sim 28.5$ keV, corresponding to a 
magnetic field of $\sim 2.5\times 10^{12}(1+z)$ G. Between November 2004 and February 2005 a giant
outburst was monitored with the Rossi XTE and Integral 
observatories in the X-ray band  \citep{mowlavi06,tsygankov06,tsygankov10}. The
position of the peak energy of the cyclotron line changed during
the outburst in anti-correlation with the luminosity, going from
$\sim 30$ keV at the onset of the outburst to $\sim 24$ keV at the luminosity
peak, and returning to $\sim 30$ keV at the end of the outburst, following the
same track in both the brightening and the fading phase. This is 
interpreted with the infalling plasma 
being decelerated along the accretion columns by a  radiation dominated 
shock \citep{basko76, burnard91, becker12}, whose height
above the NS surface increases with luminosity. If the height of the line 
forming region follows the height of the shock, the observed 
anti-correlation is straightforwardly explained as the magnetic field weakens
with the distance from the polar cap.

This Letter is focused on the study of the CRSF evolution during the bright 
outburst of V0332+53 occurred in 2015. 
 Section 2 describes the BAT and XRT data 
reduction. Section 3  describes the broad 
band spectral analysis.  In Sect.\ 4 we discuss
our results.

	\section{Data reduction\label{data}}
V0332+53 went into outburst between June and September 2015.
The Burst Alert Telescope (BAT, \citealp{barthelmy05}) on board the 
Swift observatory \citep{gehrels04} performed a nearly continuous monitoring in the 
15-150 keV energy band, while the Swift
X-Ray Telescope (XRT, \citealp{burrows05}) covered the soft X-ray band 
(0.6-10 keV) with several pointed observations. 
BAT is a coded mask telescope that observes the sky in the 15--150 keV energy 
range with a field of view of 1.4 steradian (half coded). It is devoted
to an all-sky monitoring with the main aim of capturing Gamma Ray Burst events. 
The pointing strategy of Swift, that performs frequent slews to observe 
different sky directions, allows BAT to monitor more than $80\%$ of the entire 
sky every day. 
V0332+53 was observed by BAT with a daily duty-cycle of $\sim 20\%$.
The BAT survey data stored in the HEASARC public
archive\footnote{http://heasarc.gsfc.nasa.gov/docs/archive.html} were
processed with the  {\sc bat\_imager} code \citep{segreto10}, a software built for 
the analysis of data from coded mask instruments that performs screening, 
mosaicking and source detection. The background subtracted light curve and 
spectra of V0332+53 were produced with the maximum time resolution allowed by 
the data (typically 300 s). To obtain spectra with similar and sufficient
statistics, we selected time intervals of different duration to
achieve a minimum signal to noise ratio (SNR) of 300. Only for the first and the last
interval, in order to measure $E_{cyc}$ at the lowest fluxes,
the spectra were cumulated with a lower SNR ($\sim 150$). With this selection we
obtained 22 spectra.
We used the official BAT spectral redistribution
matrix\footnote{http://heasarc.gsfc.nasa.gov/docs/heasarc/caldb/data/swift/bat/index.html}.
XRT (0.2-10 keV) observed V0332+53 thirty-one times 
during the outburst.
For each BAT spectral time interval we used only the temporally closest XRT 
observation. All the selected observations are in Windowed Timing
mode \citep{hill04}.
The XRT data were processed with standard 
filtering and screening criteria ({\sc xrtpipeline v.0.12.4}, Heasoft 
v. 6.12). For each observation  we extracted the spectrum by selecting data 
 from a rectangular region of 40 pixel side along the image strip (1
pixel = 2.36'') centered on the source brightest pixel; 
the background was extracted from a region of the same size far from the
source extraction region.
The spectra were re-binned with a minimum of 20 counts per energy channel to
allow for the use of $\chi^2$ statistics. 
The XRT ancillary response files were generated with 
{\sc xrtmkarf}\footnote{http://heasarc.gsfc.nasa.gov/ftools/caldb/help/xrtmkarf.html}. 
We used the spectral redistribution matrix v014; the spectral analysis was 
performed using XSPEC v.12.5. 
Errors are reported at 68\,\% confidence level.
The luminosity of the source has been evaluated using the  1-150 keV flux 
derived from the best fit model of the continuum and assuming isotropic
emission at a distance of 7 kpc \citep{negueruela99}.

\begin{figure}
\begin{center}
\includegraphics[width=9cm]{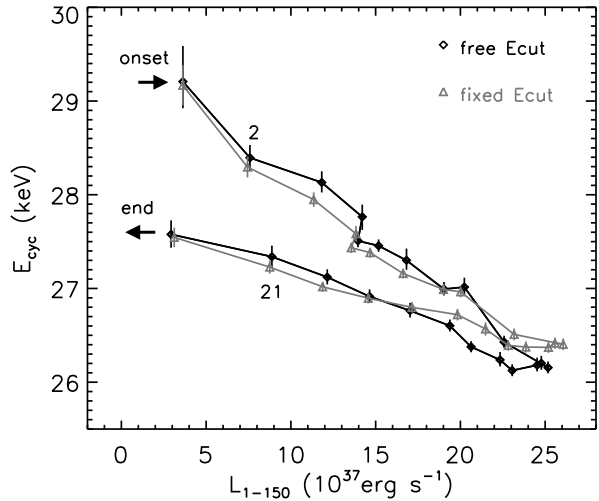}

        \end{center}
\caption{Position of the fundamental cyclotron line as a function of
luminosity. The line connecting the points indicates their temporal sequence. The
black diamonds are obtained fitting the spectra with E$_{cut}$ as a free parameter.
The grey triangles are obtained fixing E$_{cut}$ to an avarage value of 6.65 keV.
}               
		\label{evsflux} 
        \end{figure}

\section{Data analysis\label{xrt}}
Figure~\ref{lc} shows the 1--150 keV luminosity ($L_{1-150}$)
of the 2015 outburst  derived from the BAT monitoring.
Its shape can be described with a linear rise and decline,
roughly specular with respect to a peak that lasted 
$\sim 10$ days  reaching a luminosity of $\sim 2.5 \times 10^{38}$ erg s$^{-1}$.

To cope with the non-simultaneity of
the BAT and XRT data, we included in the spectral model a multiplicative
factor that takes into account any mismatch in intensity between the hard and
soft data. 
To model the continuum emission of the source we tested different spectral
shapes, modified at low energy by photo-electric absorption along the line 
of sight: a simple power law modified with an exponential cut-off, a
negative-positive power laws with exponential cutoff (NPEX \citealp{mihara95}), 
an optically thick Comptonization ({\tt comptt} in {\sc xspec}). 
We verified that the choice of the 
continuum model does not affect the results discussed in this Letter. 
In the following we will adopt a continuum described by a power law 
modified with an 
exponential cut-off ({\tt cutoffpl} in XSPEC).
We included in the model two CRSFs, the fundamental line and its second 
harmonic, using absorption Gaussian profiles ({\tt gabs} in XSPEC).
The residuals in all the 22 spectra do not statistically require any additional
component to model the fundamental CRSF.
The results of the spectral analysis 
are reported in Table 1. The continuum emission does not show any noteworthy 
variability:  the photon index does not change significantly along the outburst, while
the cutoff energy ranges between 6.2 and 7.1 keV. 
The position of the fundamental  (Fig.~\ref{lc}), the line width 
and its depth
are determined on average with relative uncertainties of $\sim 0.3\%$,  $2.8\%$, and
$4.3\%$, respectively. The relative uncertainties on the parameters of the second 
harmonic (detected at $\sim 50$ keV)
are much higher, and in the following we will focus only on the energy of the 
fundamental line (E$_{cyc}$).  Figure~\ref{spec2_21} shows the data, 
best fit model and residuals for two
representative spectra with similar luminosity in the rise and decline
(spectra 2 and 21 in Table~\ref{fit}).

The cyclotron energy shows
a clear anticorrelation with luminosity (Fig.~\ref{evsflux}, black diamonds).
Moreover, the cyclotron energy describes clearly two 
different diverging tracks for the outburst rise and decline, 
being systematically lower 
after the peak, reaching a final value of $27.68\pm 0.15$ keV versus an initial 
value of $29.2^{+0.4}_{-0.3}$ keV. 
We investigated the correlation between E$_{cyc}$ and the other fit
parameters. Figure~\ref{contour} 
shows the confidence contour plots between the energy
of the fundamental and the photon index, the cutoff energy and the width of the 
line, respectively, evaluated for the two representative spectra 2 and 21. The 
plots show that there is
no significant correlation between E$_{cyc}$ and the continuum parameters,
and a weak correlation between E$_{\rm cyc}$ and the line width $\sigma$.
The single-parameter 68\% error ranges evaluated from the fit procedure
represent adequately  the bi-dimensional confidence contours. Moreover, for
each couple of parameters, the contour plots relevant to the two spectra are
very well detached. 
We have also checked if the variability path we observe in the line
energy is biased by the variation of E$_{cut}$ measured along the
outburst. Therefore, we have fitted the spectra fixing  E$_{cut}$ to the average 
of the values reported in Table~\ref{fit}, obtaining again a double track that
overlaps with excellent agreement the one obtained when  E$_{cut}$ is left as a
free parameter (Fig.~\ref{evsflux}, grey triangles).
Finally, we note that cyclotron lines may have 
complex profiles \citep{schonherr07,nishimura13}, as also reported by 
some authors for V0332+53 \citep{pottschmidt05,nakajima10}.
The observational limits in our data do not allow us to assess 
if a change in the line profile could affect 
the determination of the centroid energy, although we argue that, if present,
such a systematic would hardly produce systematically different line centroids 
at equal luminosities as shown in Fig.~\ref{evsflux}.

\begin{table*}
\renewcommand{\arraystretch}{1.8}
\begin{center}
\scriptsize
\begin{tabular}{r l l l l l l l l l l l l l}
\hline
&Time& $N_{\rm H}$       &$\Gamma$&$E_{\rm cut}$&$N_{\rm cpl}$    &$E_{\rm cyc}$&$\sigma_{\rm cyc}$&$D_{\rm cyc}$&$E_{\rm 2}$&$\sigma_{\rm 2}$&$D_{\rm 2}$&L$_{1-150}$&$\chi_{\rm red}^2$(dof) \\
\hline
 1 &	5.00& 0.79 $_{-0.10}^{+0.11}$& -0.08 $_{-0.06}^{+0.09}$& 6.7  $_{-0.3 }^{+0.7}$& 0.07 $_{-0.01}^{+0.01}$& 29.2  $_{-0.3} ^{+0.4} $& 3.6  $_{-0.3} ^{+0.4} $& 1.2  $_{-0.2} ^{+0.3} $& 50	  $_{-2}  ^{+3}  $& 3.8 $_{-1.6}^{+2.2}$& 1.3 $_{-0.8}^{+1.4}$  &     3.6 & 1.07(494)\\
 2 &   13.38& 0.81 $_{-0.09}^{+0.10}$& -0.06 $_{-0.05}^{+0.04}$& 6.91 $_{-0.13}^{+0.14}$& 0.15 $_{-0.01}^{+0.01}$& 28.39 $_{-0.14}^{+0.14}$& 3.77 $_{-0.14}^{+0.14}$& 1.25 $_{-0.07}^{+0.08}$& 48.9 $_{-0.6}^{+1.0}$& 3.6 $_{-0.7}^{+0.9}$& 0.9 $_{-0.3}^{+0.4}$        &     7.6 & 1.09(494)\\
 3 &   20.18& 0.85 $_{-0.10}^{+0.10}$& -0.03 $_{-0.05}^{+0.05}$& 7.12 $_{-0.16}^{+0.20}$& 0.23 $_{-0.02}^{+0.02}$& 28.13 $_{-0.11}^{+0.12}$& 4.08 $_{-0.12}^{+0.13}$& 1.32 $_{-0.07}^{+0.09}$& 48.6 $_{-0.7}^{+0.9}$& 3.4 $_{-0.6}^{+0.8}$& 1.6 $_{-0.4}^{+0.6}$        &     11.8& 1.14(494)\\
 4 &   24.58& 0.84 $_{-0.10}^{+0.10}$& -0.03 $_{-0.06}^{+0.06}$& 7.10 $_{-0.29}^{+0.41}$& 0.28 $_{-0.02}^{+0.02}$& 27.76 $_{-0.14}^{+0.13}$& 4.19 $_{-0.16}^{+0.17}$& 1.31 $_{-0.11}^{+0.13}$& 52	$_{-2}  ^{+2}  $& 5.9 $_{-1.3}^{+1.4}$& 1.6 $_{-0.6}^{+0.9}$    &     14.2& 1.12(494)\\
 5 &   27.05& 0.82 $_{-0.10}^{+0.10}$& -0.05 $_{-0.05}^{+0.05}$& 6.92 $_{-0.12}^{+0.14}$& 0.28 $_{-0.02}^{+0.02}$& 27.51 $_{-0.07}^{+0.08}$& 3.81 $_{-0.09}^{+0.10}$& 1.23 $_{-0.05}^{+0.06}$& 48.6 $_{-0.4}^{+0.6}$& 3.5 $_{-0.5}^{+0.6}$& 1.3 $_{-0.2}^{+0.3}$        &     14.0& 1.15(493)\\
 6 &   29.75& 0.82 $_{-0.10}^{+0.10}$& -0.05 $_{-0.05}^{+0.05}$& 6.95 $_{-0.12}^{+0.13}$& 0.30 $_{-0.02}^{+0.02}$& 27.46 $_{-0.06}^{+0.07}$& 3.81 $_{-0.09}^{+0.09}$& 1.28 $_{-0.05}^{+0.05}$& 49.3 $_{-0.5}^{+0.6}$& 3.6 $_{-0.4}^{+0.5}$& 1.6 $_{-0.3}^{+0.3}$        &     15.2& 1.10(493)\\
 7 &   32.17& 0.89 $_{-0.10}^{+0.10}$& -0.04 $_{-0.05}^{+0.05}$& 6.87 $_{-0.10}^{+0.11}$& 0.35 $_{-0.02}^{+0.03}$& 27.20 $_{-0.06}^{+0.06}$& 3.81 $_{-0.08}^{+0.08}$& 1.29 $_{-0.04}^{+0.05}$& 48.7 $_{-0.5}^{+0.5}$& 3.9 $_{-0.4}^{+0.5}$& 1.3 $_{-0.2}^{+0.2}$        &     16.8& 1.10(510)\\
 8 &   35.23& 0.82 $_{-0.10}^{+0.10}$& -0.07 $_{-0.05}^{+0.05}$& 6.66 $_{-0.11}^{+0.12}$& 0.39 $_{-0.03}^{+0.03}$& 26.99 $_{-0.07}^{+0.08}$& 4.10 $_{-0.08}^{+0.09}$& 1.26 $_{-0.04}^{+0.05}$& 49.5 $_{-0.6}^{+0.8}$& 3.9 $_{-0.6}^{+0.7}$& 1.3 $_{-0.3}^{+0.3}$        &     19.0& 1.12(494)\\
 9 &   38.68& 0.82 $_{-0.10}^{+0.10}$& -0.07 $_{-0.05}^{+0.06}$& 6.78 $_{-0.16}^{+0.23}$& 0.40 $_{-0.03}^{+0.03}$& 27.01 $_{-0.09}^{+0.10}$& 4.21 $_{-0.10}^{+0.12}$& 1.32 $_{-0.07}^{+0.09}$& 49.5 $_{-0.7}^{+0.7}$& 4.8 $_{-0.8}^{+0.9}$& 1.2 $_{-0.3}^{+0.4}$        &     20.2& 1.10(494)\\
10 &   41.97& 0.78 $_{-0.10}^{+0.10}$& -0.11 $_{-0.05}^{+0.05}$& 6.42 $_{-0.10}^{+0.11}$& 0.45 $_{-0.03}^{+0.04}$& 26.43 $_{-0.06}^{+0.07}$& 4.21 $_{-0.08}^{+0.09}$& 1.26 $_{-0.04}^{+0.05}$& 47.8 $_{-0.5}^{+0.6}$& 3.4 $_{-0.4}^{+0.5}$& 1.2 $_{-0.2}^{+0.3}$        &     22.6& 1.13(494)\\
11 &   46.13& 0.76 $_{-0.10}^{+0.10}$& -0.14 $_{-0.05}^{+0.05}$& 6.21 $_{-0.09}^{+0.10}$& 0.51 $_{-0.04}^{+0.04}$& 26.16 $_{-0.06}^{+0.07}$& 4.31 $_{-0.08}^{+0.09}$& 1.18 $_{-0.04}^{+0.05}$& 48.3 $_{-0.6}^{+0.6}$& 3.1 $_{-0.5}^{+0.7}$& 1.2 $_{-0.3}^{+0.3}$        &     25.2& 1.15(494)\\
12 &   50.05& 0.75 $_{-0.10}^{+0.10}$& -0.14 $_{-0.05}^{+0.05}$& 6.21 $_{-0.10}^{+0.11}$& 0.49 $_{-0.04}^{+0.04}$& 26.20 $_{-0.07}^{+0.08}$& 4.37 $_{-0.09}^{+0.10}$& 1.19 $_{-0.05}^{+0.06}$& 47.6 $_{-0.6}^{+0.8}$& 3.4 $_{-0.6}^{+0.8}$& 0.9 $_{-0.2}^{+0.3}$        &     24.8& 1.16(494)\\
13 &   52.71& 0.75 $_{-0.10}^{+0.10}$& -0.14 $_{-0.05}^{+0.05}$& 6.23 $_{-0.10}^{+0.12}$& 0.49 $_{-0.04}^{+0.04}$& 26.18 $_{-0.07}^{+0.08}$& 4.27 $_{-0.09}^{+0.10}$& 1.18 $_{-0.04}^{+0.06}$& 48.7 $_{-0.9}^{+1.1}$& 3.8 $_{-0.8}^{+1.1}$& 0.9 $_{-0.3}^{+0.4}$        &     24.5& 1.15(494)\\
14 &   55.54& 0.75 $_{-0.10}^{+0.10}$& -0.15 $_{-0.05}^{+0.05}$& 6.14 $_{-0.10}^{+0.10}$& 0.46 $_{-0.03}^{+0.04}$& 26.13 $_{-0.06}^{+0.07}$& 4.19 $_{-0.08}^{+0.09}$& 1.14 $_{-0.04}^{+0.04}$& 49.2 $_{-0.7}^{+0.8}$& 3.5 $_{-0.6}^{+0.7}$& 1.4 $_{-0.3}^{+0.4}$        &     23.1& 1.14(494)\\
15 &   58.36& 0.75 $_{-0.10}^{+0.10}$& -0.14 $_{-0.05}^{+0.05}$& 6.26 $_{-0.11}^{+0.14}$& 0.44 $_{-0.03}^{+0.03}$& 26.24 $_{-0.07}^{+0.09}$& 4.24 $_{-0.09}^{+0.10}$& 1.19 $_{-0.05}^{+0.06}$& 51	$_{-1}  ^{+2}  $& 4.5 $_{-0.9}^{+1.4}$& 1.3 $_{-0.4}^{+0.7}$    &     22.3& 1.11(494)\\
16 &   61.05& 0.75 $_{-0.10}^{+0.10}$& -0.15 $_{-0.05}^{+0.05}$& 6.22 $_{-0.10}^{+0.10}$& 0.41 $_{-0.03}^{+0.03}$& 26.38 $_{-0.06}^{+0.06}$& 4.12 $_{-0.07}^{+0.08}$& 1.19 $_{-0.04}^{+0.04}$& 48.2 $_{-0.6}^{+0.7}$& 3.0 $_{-0.5}^{+0.5}$& 1.4 $_{-0.3}^{+0.3}$        &     20.6& 1.15(494)\\
17 &   63.69& 0.77 $_{-0.10}^{+0.10}$& -0.12 $_{-0.05}^{+0.05}$& 6.37 $_{-0.10}^{+0.11}$& 0.39 $_{-0.03}^{+0.03}$& 26.60 $_{-0.06}^{+0.07}$& 4.10 $_{-0.07}^{+0.08}$& 1.24 $_{-0.04}^{+0.05}$& 48.5 $_{-0.8}^{+1.0}$& 3.5 $_{-0.6}^{+0.8}$& 1.0 $_{-0.2}^{+0.3}$        &     19.4& 1.12(494)\\
18 &   68.12& 0.79 $_{-0.10}^{+0.10}$& -0.10 $_{-0.05}^{+0.05}$& 6.56 $_{-0.13}^{+0.17}$& 0.34 $_{-0.02}^{+0.03}$& 26.76 $_{-0.07}^{+0.08}$& 3.99 $_{-0.08}^{+0.10}$& 1.29 $_{-0.05}^{+0.07}$& 49.5 $_{-0.8}^{+1.0}$& 4.7 $_{-0.7}^{+0.9}$& 1.3 $_{-0.3}^{+0.4}$        &     17.0& 1.14(494)\\
19 &   74.28& 0.81 $_{-0.10}^{+0.10}$& -0.07 $_{-0.05}^{+0.05}$& 6.71 $_{-0.13}^{+0.16}$& 0.29 $_{-0.02}^{+0.02}$& 26.91 $_{-0.07}^{+0.08}$& 3.92 $_{-0.08}^{+0.09}$& 1.28 $_{-0.05}^{+0.06}$& 48.7 $_{-0.8}^{+0.9}$& 4.3 $_{-0.7}^{+0.8}$& 1.1 $_{-0.3}^{+0.4}$        &     14.6& 1.15(494)\\
20 &   81.13& 0.84 $_{-0.10}^{+0.10}$& -0.04 $_{-0.06}^{+0.06}$& 6.98 $_{-0.18}^{+0.23}$& 0.25 $_{-0.02}^{+0.02}$& 27.12 $_{-0.08}^{+0.08}$& 4.00 $_{-0.09}^{+0.10}$& 1.34 $_{-0.06}^{+0.08}$& 49.4 $_{-0.8}^{+0.9}$& 5.3 $_{-0.7}^{+0.8}$& 1.3 $_{-0.3}^{+0.4}$        &     12.1& 1.13(494)\\
21 &   88.86& 0.82 $_{-0.10}^{+0.10}$& -0.05 $_{-0.06}^{+0.06}$& 6.93 $_{-0.24}^{+0.34}$& 0.18 $_{-0.01}^{+0.01}$& 27.34 $_{-0.12}^{+0.12}$& 4.00 $_{-0.14}^{+0.15}$& 1.32 $_{-0.09}^{+0.11}$& 48.6 $_{-1.0}^{+1.0}$& 4.9 $_{-1.1}^{+1.2}$& 1.2 $_{-0.4}^{+0.6}$        &     8.9 & 1.13(494)\\
22 &  105.04& 0.81 $_{-0.10}^{+0.10}$& -0.05 $_{-0.06}^{+0.07}$& 7.03 $_{-0.31}^{+0.61}$& 0.06 $_{-0.01}^{+0.01}$& 27.68 $_{-0.15}^{+0.15}$& 3.71 $_{-0.17}^{+0.19}$& 1.34 $_{-0.12}^{+0.16}$& 47	$_{-1}  ^{+2}  $& 4.6 $_{-0.9}^{+1.8}$& 1.4 $_{-0.5}^{+1.2}$    &     2.9 & 1.12(494)\\
\hline

\end{tabular}
\caption{Best fit spectral parameters for each BAT+XRT broad band
spectrum.
The second column lists the central time of each spectrum in days, referred to 
MJD 57188.24 (the start time of the first spectrum)
The broad band X-ray model used to fit the continuum is a power-law with 
photon index $\Gamma$ and normalization $N_{cpl}$ in photons 
kev$^{-1}$cm$^{-2}$s$^{-1}$ at 1 keV,
modifield by  an exponential cutoff at energy $E_{cut}$ (keV). The model
includes also photo-electric absortion by neutral inter-stellar matter  
($N_{\rm H}$, in units of $10^{22}$ atoms cm$^{-2}$).
The CRSFs are  modeled with two Gaussians in absorption for the
fundamental and the second harmonic, respectively. The central energies
$E_{\rm cyc}$ and $E_2$ and the line widths $\sigma_{\rm cyc}$ and $\sigma_2$ 
are in keV. $D_{\rm cyc}$ and $D_2$ are the optical depths 
of the features.
The luminosity in the 1-150 keV range is in units of $10^{37}$ erg s$^{-1}$
}
 \label{fit}
\end{center}
\end{table*}

\begin{figure}
\begin{center}
\includegraphics[width=6cm,angle=270]{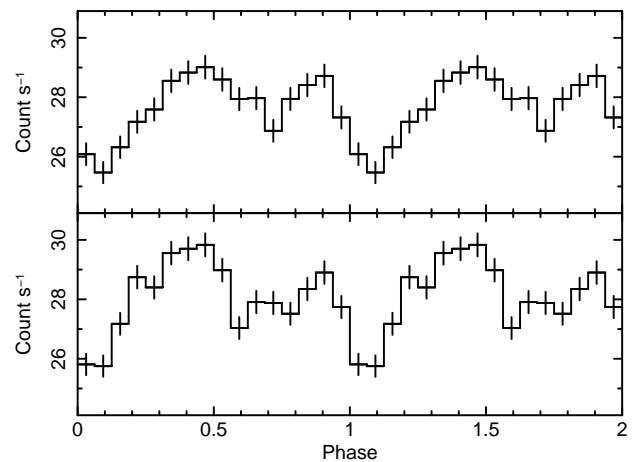}

        \end{center}
\caption{Pulse profiles obtained using the XRT data
close to BAT intervals 2
(XRT ObsID 00031293007, 00031293008, 00031293009, top panel) and 
21 (XRT ObsID 00081588002, 00081588003, bottom panel)}               
\label{profili}
        \end{figure}

\section{Discussion\label{discuss}}
%
The anticorrelation between the fundamental CRSF energy and luminosity was already 
observed in the 2004/05 
outburst \citep{tsygankov10}. However, in the 2015 outburst we find a remarkably 
significant 
difference in the path described by the cyclotron energy versus luminosity:
while in the 2004/05 outburst the energy of the fundamental appears to 
follow the same path both 
during the rise and the decline, so that the intensity of the magnetic field is the
same at the onset and at the end of the outburst, in 2015 the energy of
the fundamental follows two distinct tracks, with a difference of $\sim$~1.5 keV
between the energy measured at the onset and at the end of the outburst. 
\citet{lutovinov15} have shown that the energy of the fundamental is
variable with the pulse phase and that the variability pattern changes for 
different luminosity levels.
They explain this behaviour with a change in the beam pattern
along the line of sight  
and in the structure of the emission region. The interpulse variability pattern 
may affect the position  of the CRSF fundamental in the phase-averaged spectra, 
and if the double track in  Fig.~\ref{evsflux} arises from 
different variability patterns in E$_{cyc}$ at equal
luminosity levels, we would expect a corresponding change in  
the pulse profiles.
We have produced 0.3-10 keV pulse profiles using the XRT data collected 
along the outburst: the profile shows a substantial shape evolution, from a
double peak profile at low luminosity to a single peak at high luminosity (see
also \citealp{tsygankov06}).
Figure~\ref{profili} shows that the profiles at similar luminosity levels close
to the beginning and the end of the outburst have consistent shapes.
Indeed, while both cyclotron line energy and pulse profile shape are known to 
change significantly with pulse phase and luminosity (Tsygankov 2006, 
Lutovinov 2015), a significant fraction of the energy flux is emitted in the 
soft band where, as we demonstrated, the pulse profile remains stable.
Therefore, the comparison of the
XRT profiles in the soft X-rays provides a hint against the hypothesis of a
geometrical beam variation.

If, on the other hand, the line forming region is 
the same at equal luminosities the observed difference in 
the cyclotron energy 
corresponds to a difference in the magnetic field of $\sim 1.7\times10^{11}$ G 
(assuming $z$\,=\,0.3 at the surface of a NS with a mass of 1.4\,$M_{\odot}$ and a
radius of 10 km).  

\begin{figure}
\begin{center}
\includegraphics[width=9cm]{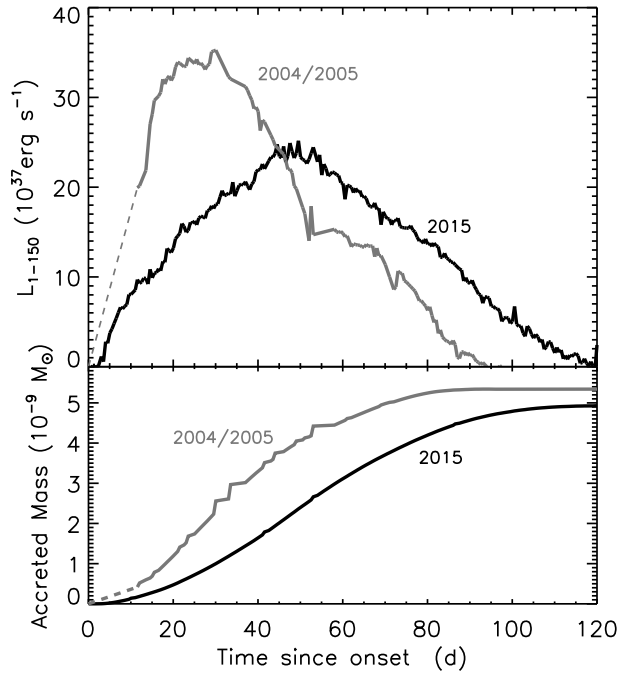}

        \end{center}
\caption{Top panel: 
light curve of the 2015 (black line) and of the 2004/05 (grey line)
outbursts, reported to the same onset time. For the first 15 days of the 2004/05 
outburst (not observed by BAT) we assume a linear rise (dashed grey line) as 
suggested by the RXTE  All-Sky Monitor light curve.
Bottom panel: total accreted mass as a function of time, according to 
$L_{1-150}=\rm \eta \dot M c^2$, where we adopt a
mass-energy conversion factor $\eta=0.15$, as commonly assumed for a 
neutron star.
}               
\label{compare}
        \end{figure}

An intriguing question is why the field drop has not been observed  
also in the 2004/05 outburst. Figure~\ref{compare}, that compares the light curves of 
the two outbursts (top panel) and the cumulative accreted mass
(bottom panel), shows a substantial difference:
although the total mass accreted at the end of the two outbursts is similar, 
during the 2004/05 outburst a higher luminosity (1.5 times that of 
2015) is reached earlier. 
Therefore, we argue that the splitting in two branches of the $E_{\rm cyc}$ 
versus  $L_{\rm 1-150}$  depends on how the mass accretes along the outburst,
and not simply on the total accreted mass, in agreeement with
previous suggestions that the decay of the magnetic
field is not directly proportional to the total accreted mass \citep{wijers97}.


Among the several mechanism that can induce a decay of the magnetic field
through accretion (see e.g. \citealp{cumming04,ruderman91,lovelace05}) the 
one that seems to accord better with the observations is diamagnetic 
screening \citep{choudhuri02}. In this hypothesis the accreting plasma 
builds up to form a magnetically confined mound, where the gas pressure 
balances the magnetic stresses. This would produce, as an overall
effect, a distortion of the field lines \citep{brown98} observed as a
decrease of the field component along the accretion column. However, 
for higher peak luminosities, the magnetic cap surface is larger and it is 
conceivable that the field at its border is weaker, preventing the gas
confinement. The fast rise time observed in the 2004/05 outburst 
could lead to this configuration at the very beginning of the outburst, 
hampering the formation of the mound.
Alternatively, the mound may have formed at an early stage of the outburst, 
reaching in a short time
the maximum size for a stable structure. After that, an 
equilibrium was reached where
the plasma settling on the mound was balanced by the matter leaking out from the
mound. In this hypothesis
we expect that for most of the outburst evolution the tracks do not diverge. This
hypothesis is suggested by the first two points of the outburst rise in Figure 3 in 
\citet{tsygankov10}, that  show E$_{cyc}$ values significantly
higher than the values at equal luminosity during the decline. However, the lack
of coverage  in the first ten days of the outburst prevents a firm conclusion on
this point.

\section*{Acknowledgments}
This work has been supported
by ASI grant I/011/07/0.

\bsp

\label{lastpage}

\end{document}